\documentclass[aps,twocolumn]{revtex4}
\usepackage{graphicx}
\newcommand {\bea}{\begin{eqnarray}}
\newcommand {\eea}{\end{eqnarray}}
\newcommand {\be}{\begin{equation}}
\newcommand {\ee}{\end{equation}}

\begin{document}


\title{QCD and the $\eta'$ Mass: Instantons 
or Confinement?}

\author{T.~Sch\"afer$^{1,2}$}

\affiliation{
$^1$Department of Physics, SUNY Stony Brook,
Stony Brook, NY 11794\\ 
$^2$Riken-BNL Research Center, Brookhaven National 
Laboratory, Upton, NY 11973}

\begin{abstract}
  We argue that lattice calculations of the $\eta'$ mass 
in QCD with $N_c=2$ colors performed at non-zero baryon chemical 
potential can be used to study the mechanism responsible 
for the mass of the $\eta'$. QCD with two colors is an ideal 
laboratory because it exhibits confinement, chiral symmetry 
breaking and a would-be $U(1)_A$ Goldstone boson at all 
densities. Since the instanton density and the confinement 
scale vary with density in a very different way, instantons
are clearly distinguishable from other possible mechanisms.
There is an instanton prediction for the $\eta'$ mass at large 
density that can be compared to lattice results. The density 
dependence of the instanton contribution is a simple 
consequence of the integer topological charge carried by the 
instanton. We also argue that $N_c=3$ color QCD at finite isospin 
density can be used in order to study the origin of OZI-violation 
in the scalar sector. 
 
\end{abstract}

\maketitle
\newpage

\section{Introduction}
\label{sec_intro}

  The $U(1)_A$ puzzle in QCD is related to the absence of 
a ninth Goldstone boson connected to the spontaneous
breakdown of the $U(1)_A$ chiral symmetry \cite{Weinberg:ui}. 
It was realized soon after the discovery of QCD that the $U(1)_A$ 
symmetry of the QCD lagrangian is anomalous, but it was also 
noted that the divergence of the $U(1)_A$ current is itself
a total divergence \cite{Fritzsch:pi}. Superficially, it 
would then seem that the $U(1)_A$ anomaly is not sufficient
to remove the $U(1)_A$ Goldstone boson. The puzzle was resolved 
after Belavin, Polyakov, Schwartz and Tyupkin discovered 
topological structures, instantons, in QCD \cite{Belavin:fg}.
't Hooft showed that instantons lead to the violation 
of axial charge \cite{'tHooft:up}, and that instantons 
induce an effective $(2N_f)$-fermion operator which 
contributes to the $\eta'$ meson mass \cite{'tHooft:up,'tHooft:1986nc}.

  Since then, lattice QCD calculations have identified 
instantons, verified the presence of fermion zero modes, 
and established their relation to the mass of the $\eta'$ 
\cite{Edwards:2001ei}. Also, the instanton liquid model was 
expanded into a phenomenologically successful description 
of chiral symmetry breaking and the $U(1)_A$ anomaly 
\cite{Schafer:1996wv,Diakonov:1995ea}. Unfortunately, we 
cannot compute the instanton contribution to the $\eta'$ 
mass from first principles. As a consequence, there are 
still speculations that the $\eta'$ mass is related to 
structures with fractional topological charge that do not 
appear in the classical limit, or that the $\eta'$ mass 
is in some way related to confinement \cite{Isgur:2000ts}. 
The latter suggestion was first made by Kogut and Susskind 
prior to the discovery of instantons in QCD \cite{Kogut:ab}.

  Even if we cannot do a parameter-free calculation of 
the $\eta'$ mass in QCD we can still  try to distinguish 
different mechanisms for generating the $\eta'$ mass by 
their scaling behavior. In QCD, of course, any contribution 
to the $\eta'$ mass has to be proportional to the QCD scale 
parameter $\Lambda_{QCD}$. Witten suggested that the number 
of colors $N_c$ could be used as a parameter \cite{Witten:1978bc}. 
He argued that the instanton contribution scales as $\exp(-N_c)$
whereas effects related to confinement give $m_{\eta'}^2\sim 
1/N_c$. However, the relation $m_{\eta'}^2\sim\exp(-N_c)$ is
only correct for very small instantons, $\rho\ll
\Lambda_{QCD}^{-1}$. Indeed, we recently argued that 
the instanton contribution to the $\eta'$ mass also
scales as $m_{\eta'}^2\sim 1/N_c$ \cite{Schafer:2002af}.

  Another possibility is to use the temperature $T$ as
a parameter. The density of instantons is expected
to be suppressed by a large power of $T$, $(N/V)\sim 
T^{b-4}$, where $b=11N_c/3-2N_f/3$ is the first coefficient 
of the QCD beta function 
\cite{Gross:1980br,Pisarski:ms,Alkofer:rr,Kapusta:1995ww}.
Effects related to confinement or topological objects 
other than instantons would presumably have a different
dependence on temperature. The problem with this idea is 
that the suppression of instanton effects is related to 
perturbative color screening. This implies that the power 
law suppression only applies if the temperature is larger 
than the critical temperature $T_c$ for chiral symmetry 
restoration \cite{Shuryak:1994ay}. In order to have 
rigorous theoretical control over instanton effects 
we have to consider the quark gluon plasma phase instead
of the hadronic phase.
  
  In this note we suggest using the baryon chemical 
potential $\mu$ as a parameter. We shall argue that this 
parameter is more useful than the number of colors or 
the temperature because there is chiral symmetry breaking 
and a hadronic phase for all values of $\mu$, and there 
is theoretical control over both the instanton contribution 
to the $\eta'$ mass and the scale of confinement effects. 
We first formulate our proposal in terms of $SU(2)$ gauge 
theory at finite baryon density. We then show that a similar
situation arises in $SU(3)$ gauge theory at finite
isospin density. Both of these theories can be studied 
with lattice algorithms that are available today. 

  We should note that many of the features that we 
shall discuss also apply, with some modifications, 
to $SU(3)$ gauge theory at finite baryon density
\cite{Schafer:1999ef}. However, this theory cannot
be studied on the lattice at present, and we shall
not discuss it in detail. 

\section{QCD with two colors}
\label{sec_nc2}

  Let us summarize some of the salient features of $SU(2)$
gauge theory at zero and non-zero baryon chemical potential
\cite{Peskin:1980gc,Rapp:1998zu,Kogut:1999iv,Kogut:2000ek,Splittorff:2002xn}. 
For simplicity, we will concentrate on $N_f=2$ flavors. $SU(2)$ 
gauge theory has a meson spectrum which is very similar to 
$SU(3)$ QCD. Baryons, on the other hand, are bosons rather 
than fermions and their spectrum is very different as compared 
to $N_c=3$ QCD. Because the $SU(2)$ gauge group is pseudo-real, 
there is a Pauli-G\"ursey symmetry which relates quarks and 
anti-quarks. This symmetry mixes the quark-anti-quark condensate 
$\langle\bar{q}^aq^a\rangle$ with the diquark condensate $\langle 
\epsilon^{ab}q^{a\,T}C\gamma_5\tau_2 q^b\rangle$. Here, $a,b$
are color indices and $\tau_2$ is the anti-symmetric $SU(2)_F$ 
flavor matrix. As a result the chiral symmetry group is $SU(4)$ 
rather than $SU(2)_L\times SU(2)_R$.

\begin{figure}
\centering
\includegraphics[width=7cm]{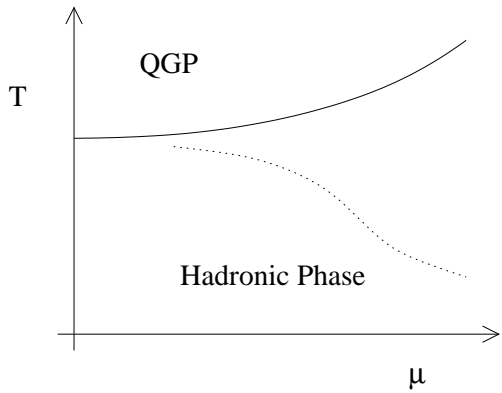}
\caption{\label{fig_nc2}
Phase structure of $N_c=2$ QCD at finite baryon chemical 
potential and temperature. We consider the diquark phase
$m=0,j\to 0$. On the $T\neq 0$ line the symmetry breaking
pattern is $SU(4)\to Sp(4)$. Above $T_c$, the $SU(4)$ 
symmetry is restored. At $\mu\neq 0$ the $SU(4)$ symmetry
is explicitly broken to $SU(2)\times SU(2)\times U(1)$.
For $T<T_c$ this symmetry is spontaneously broken to 
$SU(2)\times SU(2)$. The dashed line is the location 
of the deconfining phase transition. There is no strict
order parameter for this transition at low density
and the transition may just be a rapid cross-over.}
\end{figure}

 Let us consider chiral symmetry breaking in the presence 
of a small source term 
\be
{\cal L}_s=m\bar{q}^aq^a+\frac{j}{2}
  \left(\epsilon^{ab}q^{a\,T}C\gamma_5\tau_2 q^b+ h.c.\right).
\ee 
Chiral symmetry is broken according to $SU(4)\to Sp(4)$ in 
the low-temperature and low-density phase for both $j=0,m\to 0$ 
and $m=0,j\to 0$. In the case $m\neq 0$ the order parameter 
is the quark-anti-quark condensate $\langle\bar{q}^aq^a
\rangle$. There are 5 Goldstone bosons, three pions $\vec{\pi}$,
the scalar diquark $S$ and the scalar anti-diquark $\bar{S}$. 
Because of the $U(1)_A$ anomaly the would-be singlet Goldstone 
boson, the $\eta'$, is heavy. In the case $j\neq 0$ the order 
parameter is the diquark condensate $\langle q^TC\gamma_5\tau_2 
q\rangle$. Again there are 5 Goldstone bosons, three pions 
$\vec{\pi}$, the sigma $\sigma$ and the scalar diquark $S$. 
The would-be singlet Goldstone boson, the pseudoscalar diquark 
$P$, is heavy. 

 For $\mu=0$ all directions of the source term $(m,j)=m_0
(\cos(\alpha),\sin(\alpha))$ are physically equivalent. In 
particular, the masses of the pseudoscalar diquark $P$ in 
the diquark phase is equal to the mass of the $\eta'$ in 
quark-anti-quark condensed phase. If the baryon chemical 
potential is non-zero the $SU(4)$ symmetry is broken explicitly. 
In the following, we shall consider the diquark phase $m=0,j\to 0$. 
In this case, the phase diagram is particularly simple, see
Fig.~1. Chiral symmetry is spontaneously broken at 
$T=\mu=0$. There is a critical temperature $T_c$ such 
that chiral symmetry is restored for $T>T_c$. However, 
most likely, there is no phase transition as a function
of the chemical potential for $T<T_c$. This, of course,
is the main feature that distinguishes the $SU(2)$ theory 
at non-zero $\mu$ as a laboratory for studying the 
$U(1)_A$ anomaly. 

  The effective lagrangian for the singlet pseudoscalar 
Goldstone boson is 
\be
\label{l_nc2}
 {\cal L} = f_P^2\left[ (\partial_0\phi)^2-v^2(\partial_i\phi)^2
 \right] - V(\phi).
\ee
The decay constant and Goldstone boson velocity can be determined 
in perturbation theory. At leading order, the result is \cite{Son:2001jm}
\be
\label{f_nc2}
 f_P^2 = \left( \frac{\mu^2}{8\pi^2}\right),
 \hspace{1cm} v^2 = \frac{1}{3}.
\ee
The potential $V(\phi)$ receives contributions from instantons.
We find $V(\phi)=-A_P\cos(\phi+\theta)$ where $\theta$ is the 
QCD theta angle. If the chemical potential is big, $\mu\gg
\Lambda_{QCD}$, large instantons are suppressed and the coefficient 
$A$ can be determined in perturbation theory. We find
\cite{Son:2001jm,Schafer:2002ty} 
\be
\label{A_nc2}
 A_P = C_{2,2} 6\pi^4 \left[\frac{4\pi}{g}\Delta
     \left(\frac{\mu^2}{2\pi^2}\right)\right]^2
  \left(\frac{8\pi^2}{g^2}\right)^{4}
  \left(\frac{\Lambda}{\mu}\right)^{8}\Lambda^{-2}
\ee
with 
\be
  C_{N_c,N_f} = \frac{0.466\exp(-1.679N_c)1.34^{N_f}}
    {(N_c-1)!(N_c-2)!}.
\ee
At large $\mu$ the gap $\Delta$ can also be determined in 
perturbation theory. We get
\cite{Son:1999uk,Schafer:1999jg,Hong:2000fh,Pisarski:2000tv}
\be 
\label{gap_nc2}
 \Delta = 512\pi^4 b_0'\mu g^{-5}
 \exp\left(-\frac{2\pi^2}{g(\mu)}\right),
\ee 
where the parameter $b_0'=\exp(-(\pi^2+4)(N_c-1)/16)$ controls 
the size of non-Fermi liquid effects \cite{Brown:1999aq,Wang:2001aq}. 
Using equ.~(\ref{f_nc2}-\ref{gap_nc2}) we can determine 
the mass of the pseudoscalar Goldstone boson. We have
\be
\label{m_P}
 m_P^2 = \frac{A_P}{2f_P^2}.
\ee
The result has the structure of the Witten-Veneziano
relation where $A_P$ plays the role of the topological
susceptibility. We note, however, that $\chi_{top}=0$
as expected for a theory with massless fermions. In fact,
$A_P$ governs local, not global, fluctuations of the topological 
charge. For a dilute gas of instantons $A_P=(N/V)$ where $(N/V)$ 
is the density of instantons. This relation is exact in the 
limit of large baryon density. We also note that in the 
instanton liquid model $A_P\simeq (N/V)$ is very well 
satisfied even at zero baryon density.

\begin{figure}
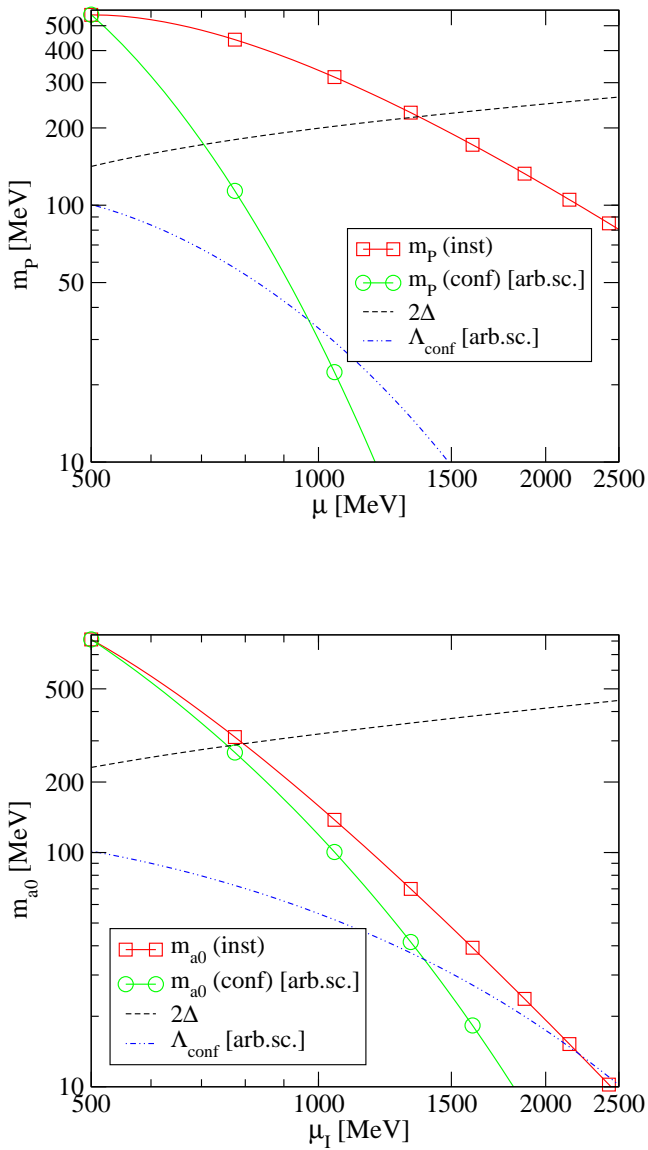

\centering
\includegraphics[width=8.5cm]{mps_nc2.eps}

\vspace*{1.4cm}
\includegraphics[width=8.5cm]{msc_mui.eps}
\caption{\label{fig_nc2_2}
Fig. a) shows the pseudoscalar Goldstone boson mass in 
$N_c=2$ QCD. The solid line marked by squares is the result 
of the instanton calculation. We have used the one-loop running 
coupling constant with $\Lambda_{QCD}=170$ MeV. The 
solid line marked by circles shows an estimate based
on the assumption $m_P^2f_P^2\sim\sigma^2$. The overall
scale of this curve is arbitrary. For comparison, we 
also show the energy gap $2\Delta$ and the confinement
scale. Again, the overall scale is arbitrary. Fig. b)
shows the scalar Goldstone boson mass in $N_c=3$ QCD
at finite isospin density.}
\end{figure}

We observe that equ.~(\ref{m_P}) implies
\be 
\label{m_P_scale}
 m_P \sim \Lambda \left(\frac{\Delta}{\Lambda}\right)
                  \left(\frac{\Lambda}{\mu}\right)^3
        \left[\log\left(\frac{\mu}{\Lambda}\right)\right]^{5/2}.
\ee
As expected, the instanton contribution to the $U(1)_A$ mass 
is suppressed at large baryon density. The power law is directly 
related to the topological charge of the instanton. Contributions 
from instantons with charge two or larger are suppressed by additional 
powers of $(\Lambda_{QCD}/\mu)$. By the same token we expect
the contribution of hypothetical objects with fractional charge
to dominate over instanton effects at large $\mu$.

 In practice, lattice calculations have to be carried out at finite 
diquark source or non-zero quark mass (or both). If the baryon 
density is small the pseudoscalar diquark and $\eta'$ meson are 
very heavy and the effect of a non-zero quark mass is small. At 
large $\mu$, however, the quark mass contribution is more important. 
If $\mu\gg\Lambda_{QCD}$ the quark mass contribution to the effective 
potential can be computed in perturbation theory. Using the 
methods described in \cite{Schafer:2001za} we find 
\be
 V_m(\phi)=-\frac{4\Delta^2}{3\pi^2}\det(M)e^{-i\phi}+h.c.,
\ee
where $M$ is the mass matrix. We observe that the topological
susceptibility does not vanish if the quark mass is non-zero, 
$\chi_{top}\sim \det(M)\Delta^2$. The contribution to the mass 
of the would-be $U(1)_A$ Goldstone boson is 
\be
 m_P^2 = \frac{32}{3}\frac{\Delta^2}{\mu^2} m_u m_d.
\ee
In the limit $\mu\to\infty$ the $m_q\neq 0$ contribution will 
eventually dominate over the instanton contribution. However, 
even for quark masses as large as $m_u=m_d= 40$ MeV the instanton 
contribution is expected to dominate for chemical potentials 
that can be achieved on the lattice.

 We now comment on possible effects related to confinement
\cite{Kogut:ab,Frank:1997ck,Alkofer:2000wg}. At large baryon 
density the gap in the fermion spectrum is much larger than 
the QCD scale parameter, $\Delta\gg\Lambda_{QCD}$. Since the diquark 
condensate is a color singlet, there is neither screening nor 
a Higgs effect operating at scales below $\Delta$. As a 
consequence, we expect the color $SU(2)$ to be confined at 
all densities. If there are effects related to confinement 
that contribute to the mass of the $\eta'$, then these 
effects should persist at all densities. 

 We do not know how to compute confinement related contributions
to the mass of the $\eta'$. However, if these effects exist then
they should be governed by the $SU(2)$ confinement scale. Rischke
et al.~observed that the pure glue theory below $\Delta$ is 
characterized by a non-trivial electric polarizability and
magnetic permeability \cite{Rischke:2000cn}. The effective 
action is 
\be 
\label{glue}
 S_{eff} = \frac{1}{2g^2}\int d^4x\left( 
  \epsilon \vec{E}^a\cdot\vec{E}^a 
  - \frac{1}{\lambda}\vec{B}^a\cdot\vec{B}^a \right),
\ee
where we have suppressed higher order terms that are 
suppressed by $(p/\Delta)$. The main feature of the 
effective action equ.~(\ref{glue}) is that the electric 
polarizability $\epsilon \simeq 1+g^2\mu^2/(18\pi^2\Delta^2)$ 
is very large in the limit $\mu\to\infty$. As a consequence, 
the $SU(2)$ pure gauge theory described by equ.~(\ref{glue}) 
is confined, but the confinement scale is exponentially small 
\cite{Rischke:2000cn,Sannino:2002re}
\be
\label{conf}
\Lambda_{\it conf} \sim \Delta \exp\left(
   -\frac{8\pi^2}{g(\mu)^2}\frac{3\sqrt{\epsilon}}{22}\right).
\ee
If the mass of the would-be $U(1)_A$ Goldstone boson is
dominated by confinement effects then we expect that $m_P^2 
f_P^2\sim \sigma^2 \sim \Lambda^4_{\it conf}$. In Fig.~2 we 
compare this scaling relation with the instanton prediction 
for the pseudoscalar Goldstone boson mass. Of course we do not 
know the constant of proportionality relating $m_P f_P$ and the 
string tension $\sigma$. In Fig.~2 we have arbitrarily scaled 
$m_P\sim \Lambda^2_{\it conf}/f_P$ to match the instanton 
prediction at $\mu\simeq 500$ MeV. We find that the instanton 
and confinement related contributions clearly scale differently. 
We also observe that the instanton contribution will always 
dominate at large $\mu$.

 We should note that the effective theory described by 
equ.~(\ref{glue}) is not covariant. As a consequence, the 
time-like and space-like string tensions are not the same. 
We can restore covariance by rescaling the time coordinate 
in equ.~(\ref{glue}) according to $x_0'= x_0/\sqrt{\epsilon}$.
This implies that the two string tensions differ by a factor
$\sqrt{\epsilon}$. We have not tried to take this effect into
account, because there is no definite theory that would 
predict the dependence of $m_P$ on the spacelike and 
timelike string tensions. We should also note that the 
estimate equ.~(\ref{conf}) of the confinement scale only has 
exponential accuracy, so that we cannot reliably predict
possible factors $\sqrt{\epsilon}$ in the pre-exponent. 

\section{QCD at finite isospin density}
\label{sec_mui}

 The large mass of the $\eta'$ implies that violations of 
the OZI rule in the pseudoscalar meson sector are substantial.
However, the $\eta'$ sector is not the only channel in which 
OZI violation is large. In particular, the OZI violating mass 
difference between the isovector-scalar $a_0$ and isoscalar-scalar
$\sigma$ meson is almost as large as the $\eta'-\pi$ splitting.
We have argued that the $a_0-\sigma$ splitting is also
dominated by instantons \cite{Schafer:1996wv,Schafer:2000hn}.
In this section we show that this idea can be checked in
$N_c=3$ QCD at finite isospin density.

 In QCD with $N_c=3$ colors and $N_f=2$ flavors chiral 
symmetry is broken according to $SU(2)_L\times SU(2)_R$. If 
the quark mass is non-zero there are three almost massless 
pions and a heavy would-be $U(1)_A$ Goldstone boson, the 
$\eta'$. There is a relatively light scalar-isoscalar meson, 
the $\sigma$, and a heavy scalar-isovector meson, the $a_0$,
which is close in mass to the $\eta'$. 

 The effect of a non-zero isospin chemical potential term
$\mu_I(u^\dagger u-d^\dagger d)$ was studied in \cite{Son:2000xc}. 
If $\mu_I>0$ the isospin chemical potential favors up quarks over 
down quarks. As a result, the mass of the positive pion is reduced. 
If $\mu_I>m_\pi/2$ pion condensation takes place and the chiral 
order parameter starts to rotate from the $\langle\bar{q}q\rangle$ 
direction to the $\langle \bar{q}i\gamma_5 \tau^- q\rangle$ direction. 
There is one exact Goldstone boson, the $\pi^+$, and two heavy pions.

 At large isospin density there is a $U(1)_A$ would-be 
Goldstone boson, the $a_0^+$. The effective lagrangian 
for the $a_0^+$ is identical to the effective lagrangian 
for the $\eta'$ in $N_c=2$ QCD, see equ.~(\ref{l_nc2}). If
$\mu_I\gg\Lambda_{QCD}$ the parameters in the effective 
lagrangian can be computed in perturbation theory. The 
decay constant and Goldstone boson velocity are given by
\be
\label{f_mui}
 f_S^2 = \left( \frac{3\mu_I^2}{16\pi^2}\right),
 \hspace{1cm} v^2 = \frac{1}{3}.
\ee
The instanton contribution to the effective potential is 
given by $A_S\cos(\phi+\theta)$ with
\be
\label{a_mui}
 A_S = C_{3,2} \frac{16\pi^4\Gamma(\frac{35}{6})}{2^{35/6}} 
   \, \Phi_S^2\,
    \left(\frac{8\pi^2}{g^2}\right)^{6}
    \left(\frac{\Lambda}{\mu_I}\right)^{35/3}\Lambda^{-2},
\ee
where the gap $M$ and superfluid density $\Phi_S$ are given by
\bea 
\label{gap_mui}
 M &=& 512\pi^4 b_0'\mu g^{-5}
 \exp\left(-\frac{3\pi^2}{2g(\mu_I)}\right),\\
 \Phi_S &=& \frac{3\pi}{g}M
     \left(\frac{\mu_I^2}{2\pi^2}\right).
\eea
The mass of the $a_0^+$ satisfies a Witten-Veneziano type relation
\be
\label{m_a}
 m_{a_0}^2 = \frac{A_S}{2f_S^2}.
\ee
Again, $A_S$ is related to the density of instantons. At 
large $\mu_I$ the $a_0^+$ mass scales as
\be 
\label{m_a_scale}
 m_{a_0} \sim \Lambda \left(\frac{M}{\Lambda}\right)
                  \left(\frac{\Lambda}{\mu_I}\right)^{29/6}
   \left[\log\left(\frac{\mu_I}{\Lambda}\right)\right]^{7/2}.
\ee
The difference in the power law suppression as compared
to equ.~(\ref{m_P_scale}) is related to the difference
between the beta functions for $N_c=2,3$. 

 Finally, we observe that the pion condensate is a color
singlet and the color $SU(3)$ is expected to be confined
at large isospin density. The confinement scale is given
by
\be
\Lambda_{\it conf} \sim \Delta \exp\left(
   -\frac{8\pi^2}{g^2(\mu)}\frac{\sqrt{\epsilon}}{11}\right).
\ee
In Fig.~2b we plot the instanton prediction for the mass
of the $a_0^+$ at large isospin density. Again, we compare
the instanton prediction to the scaling relation $m_{a_0}
\sim \sigma/f_S$. We observe that the instanton contribution
is more strongly suppressed as compared to the $N_c=2$ result. 
As noted below equ.~(\ref{m_a_scale}) this is related to the
fact that the beta function is larger. The scale $\sigma/f_S$,
on the other hand, drops off more slowly as compared to 
the $N_c=2$ case. This is a consequence of the fact that
the gap is larger and as the result the polarizability 
is smaller. As asymptotically large isospin density, however,
the $a_0$ mass is again dominated by instantons.

\section{Summary}
\label{sec_sum}

  We have argued that $SU(2)$ QCD at finite 
baryon density and $SU(3)$ QCD at finite isospin density
can be used in order to study the mechanism for generating 
the mass of the would-be $U(1)_A$ Goldstone boson in QCD. 
The main point is that there is a hadronic phase at all
densities, and that both the instanton contribution and 
the confinement scale are calculable at large chemical 
potential. Furthermore, both $SU(2)$ QCD at non-zero baryon 
chemical potential and $SU(3)$ QCD at non-zero isospin density 
can be studied on the lattice with presently available methods
\cite{Nakamura:1984uz,Kogut:2002zg}.

  We would like to mention some possible difficulties 
with our proposal. We have assumed that there is no phase
transition along the finite baryon chemical potential axis
in the phase diagram. This assumption is based on the 
observation that the phenomenologically established symmetries 
of the low density phase agree with the calculated symmetries 
of the high density phase. This means that a phase transition
is not required, but of course a transition is not forbidden
either. The question of whether or not there is a phase 
transition can be studied using lattice simulations. We should 
note that even if there is transition that separates the low 
and high-density phase, we can still study the question 
whether the mechanism for generating the mass of the 
$U(1)_A$ Goldstone boson is the same in both phases.

  We have used perturbation theory in order to compute
the gap, the instanton density, and the confinement scale 
at non-zero chemical potential. We do not know how reliable
leading order perturbation theory is for baryon chemical
potentials that can be achieved in lattice calculations.
Indeed, it was argued that the perturbative expansion for 
the gap converges very slowly \cite{Rajagopal:2000rs} and 
that the instanton contribution is very sensitive to the 
value of the QCD scale parameter \cite{Schafer:2002ty}. 
However, the power law behavior of the $U(1)_A$ Goldstone
boson mass is quite robust and a simple reflection of the 
topological charge carried by the instanton. Furthermore,
the question whether the $U(1)_A$ Goldstone boson mass 
scales with the instanton density, $m_P^2f_P^2 \sim (N/V)$, 
or the string tension, $m_P^2f_P^2\sim \sigma^2$, can be 
answered directly from the lattice data, without resort 
to perturbation theory. 

Acknowledgments: I would like to acknowledge useful
discussions with K.~Splittorff, M. Stephanov and 
J.~Verbaarschot. This work was supported in part by US DOE 
grant DE-FG-88ER40388 and a DOE OJI grant.

\end{document}